\documentstyle[12pt,psfig]{article}
\begin{document}
\begin{titlepage}
\vspace{2.5cm}
\baselineskip 24pt
\begin{center}
\large\bf{Quantum Theory and Spacelike Measurements$^1$}\\
\vspace{1.5cm}
\large{
Bernd A. Berg$^{2,3}$}\\
\vspace{0.5cm}
{berg@hep.fsu.edu}\\
\end{center}
\vspace{3cm}
\begin{center}
{\bf Abstract}\\
\end{center}

Experimentally observed violations of Bell inequalities rule out
local realistic theories. Consequently, the quantum state vector 
becomes a strong candidate for providing an objective picture of 
reality. However, such an ontological view of quantum theory faces 
difficulties when spacelike measurements on entangled states have 
to be described, because time ordering of spacelike events can change
under Lorentz-Poincar\'e transformations. In the present paper it
is shown that a necessary condition for consistency is to require 
state vector reduction on the backward light-cone. A fresh approach
to the quantum measurement problem appears feasible within such a 
framework.

theories. 

\vfill
\footnotetext[1]{{Work partially supported by the Department of 
                  Energy under contract DE-FG05-87ER40319.}}
\footnotetext[2]{{Department of Physics, The Florida State University,
                      Tallahassee, FL~32306, USA. }}
\footnotetext[3]{{Supercomputer Computations Research Institute,
                      Tallahassee, FL~32306, USA.}}
\end{titlepage}

\baselineskip 24pt


In a classical, relativistic theory all actions propagate at or below 
the speed of light. Only timelike (possibly including the limit of 
lightlike) events can influence one another. Correlations of two spacelike 
measurements $A$ and $B$ are only possible through some source $S$ in 
their joint, timelike past. A scenario of this type is depicted in
figure~1, where the signal from $S$ reaches detector $A$ at point $A1$
and detector $B$ at point $B2$. Each signal branch is supposed to propagate
locally, governed by the appropriate causal, relativistic wave equation.

\begin{figure}[tbp]
\centerline{\psfig{figure=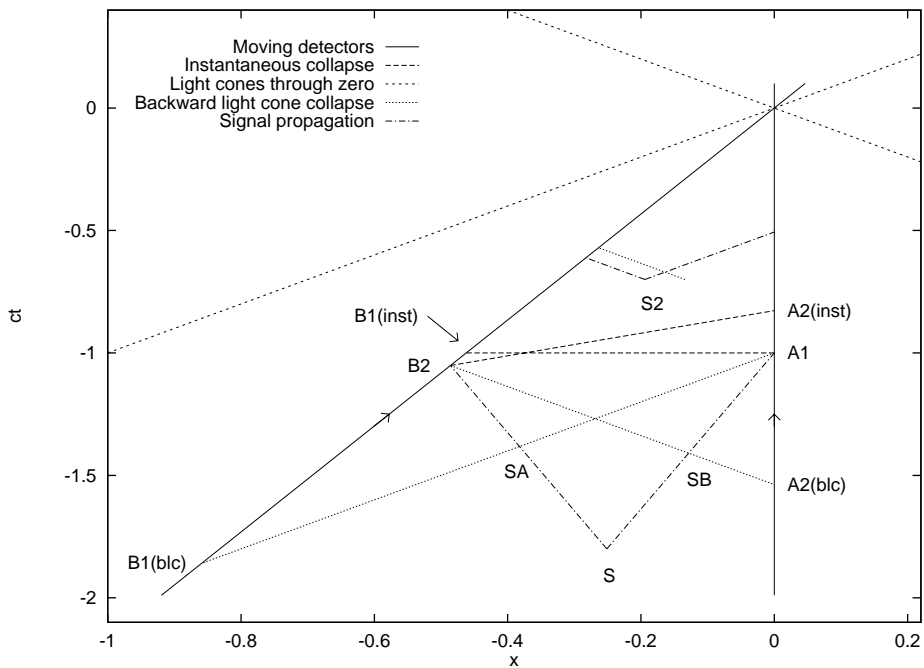}}
\caption{Illustration of the collapse problem. }
\end{figure}

Quantum correlations between such spacelike measurements are still caused
by a source in their common timelike past,
but it can be ruled out that the required information propagates at
or below the speed of light. First in this paper, we shall briefly
review this issue. Afterwards we focus on an ontological view of 
Quantum Theory (QT) in which the state vector is considered to
provide a picture of objective reality and information can propagate 
on spacelike hyperplanes through reduction (or collapse) of the state 
vector when measurements are performed. The problem of such an 
ontological understanding of QT is to find a consistent, explicit 
prescription for spacelike measurements on entangled states. It 
turns out that there is the unique implication already described in the
abstract, state vector reduction has to happen on the backward light-cone
(blc), 
which we shall discuss in some details. To have an ontological formulation
of 
QT at hand may be of importance for developing a future theory of quantum
measurement.

By now it has become textbook material, see for instance \cite{Bo93}, 
that QT implies correlations between spacelike measurements, which 
cannot be explained through local propagation at or below the speed of 
light. An often discussed arrangement is that of a singlet state which
decays into two spin 1/2 particles, whose spins are measured at
spacelike separations: with respect to direction $\hat{a}$ by 
detector $A$ and with respect to direction $\hat{b}$ by detector $B$. 
The result of each measurement is either $+1$ for spin up or $-1$
for spin down, up or down with respect to the chosen direction. In the 
following we assume that each detector has a well-defined rest-frame and, 
to begin with, that both detectors are at rest with respect to one another.

First, let both axis be the $z$-direction: $\hat{a}=\hat{b}
=\hat{e}_3$. The $+1$ measurement by detector $A$ implies the $-1$ 
measurement by
detector $B$ and vice versa. This correlation is not at all
surprising as it is classical one: When the head of a coin is send
in one direction and the the tail in the other, precisely the 
same result is achieved. Can now all measured, spacelike spin-spin 
correlation be explained this way? 

To discuss this question, let us suppose that each detector is
constructed to choose randomly, for instance with uniform probability, 
one out of a discrete number of directions $\hat{x}_i$, labelled by 
$i=1,...,n$, and then performs the spin measurement with respect to 
that axis. Hence, the delivered result is one out of $2n$ mutually 
exclusive {\it alternatives}: choice of direction and spin either 
$+1$ or $-1$ with respect to this direction. We denote these 
alternatives by
\begin{equation} \label{alternatives}
a^i,\ i=\pm 1,...,\pm n ~~{\rm and}~~ b^i,\ i=\pm 1,...,\pm n\, ,
\end{equation}
where $a^i, i\ge 1$ denotes the spin $+1$ measurement by detector $A$ 
with respect to direction $\hat{x}_i$ and $a^{-i}, i\ge 1$ the spin $-1$ 
measurement by detector $A$ with respect to direction $\hat{x}_i$. The 
$b^i$ have the same meaning, but for detector $B$. Let us introduce 
corresponding probabilities
\begin{equation} \label{probs}
p_A^i,\ i=\pm 1,...,\pm n ~~{\rm and}~~ p_B^i,\ i=\pm 1,...,\pm n\, ,
\end{equation}
where $p_A^i$ is the probability that detector $A$ picks event $a^i$ and
$p_B^i$ the probability that detector $B$ picks $b^i$. To simplify the 
arguments we neglect finite extensions of the detectors and
assume the detectors to be perfect, what means each set of 
probabilities is normalized to one: $\sum_i p_A^i=\sum_i p_B^i=1$.
Let $\vec{x}_A$ be the position of 
detector $A$ and $\vec{x}_B$ of detector $B$ with respect to their 
common rest frame. The measurement 
process is said to be spacelike when conditions are arranged as follows:
$A$ makes its measurement decision, say $a^k$, in the time interval 
$[t_A,t_A+\triangle t_A], \triangle t_A>0$ and $B$ makes its decision 
in the time interval $[t_B,t_B+\triangle t_B], \triangle t_B>0$, such 
that
\begin{equation} \label{spacelike}
c\, (\, \max\, [t_A+\triangle t_A, t_B+\triangle t_B] - 
\min\, [t_A,t_B]\, ) < | \vec{x}_B - \vec{x}_A | \, ,
\end{equation}
where $c$ is the speed at light. In words, there is no way of 
communicating 
results from $A$ to $B$ at (or below) the speed of light. Hence, an
interpretation of the spin-spin measurements as classical correlation
implies that a space-time-independent assignment of constant probabilities 
$p_A^i$, $p_B^i$ has to exist, which reproduces the observed results. 
Restrictions
on the sets of such probabilities follow from the possibility that both
detectors may pick the same direction $i\ge 1$. Then event $a^i$ measured
by $A$ implies that $b^{-i}$ is measured with {\it certainty} by $B$ or 
vice versa. Implications are, for instance,
$$ p^i_A\, p^{-i}_A = p^i_B\, p^{-i}_B = 0 ~~{\rm and}~~
p_A^i=0 \Longleftrightarrow p_B^{-i}=0 ~~{\rm for}~~~ i=\pm 1,...,\pm n\,
$$
It was shown by Bell~\cite{Bell} that such constraints lead to
inequalities which violate QT predictions. Subsequently, many experiments
were performed. They showed violations of Bell's inequalities and
confirmed 
QT. The first results, which actually achieved spacelike measurements, 
seem to be those of ref.~\cite{Aspect}. Recent experiments~\cite{Tapster} 
relate to Franson's~\cite{Fr89} realization of 
Bell's inequality.  Our emphasize is on the fact that an interpretation 
of spacelike measurements as a classical correlation between pre-defined 
probabilities (\ref{probs}) is 
excluded, see ref.~\cite{Gr89} for a particularly strong example and 
ref.~\cite{Mermin} for pedagogical reviews, which inspired some of our
notation.

It appears now natural, see for instance \cite{Penrose}, to take the
quantum state vector as substitute for the lost classical reality. To 
investigate whether 
(and in what sense) we may attribute reality to quantum states, let us
define the 
probabilities (\ref{probs}) as expectation values calculated from the QT
state 
$|\Psi_S\rangle$, which describes the system before measurement has taken
place (the subscript $S$ refers to the source and is omitted when we refer 
to the QT state after measurement). In our example
\begin{equation} \label{qmaver}
p_A^{\pm i} = \left| \langle \Psi_A^{\pm i} | \Psi_S \rangle \right|^2 
~~{\rm and}~~~
p_B^{\pm i} = \left| \langle \Psi_B^{\pm i} | \Psi_S \rangle \right|^2
~~{\rm for}~~ i=1,2,..,n\, .
\end{equation}
Here $i=1,...,n$ labels the axis and $|\Psi_A^{\pm i}\rangle$, 
$|\Psi_B^{\pm i}\rangle$ are the eigenfunctions of suitable operators 
${\cal A}^i$, ${\cal B}^i$. To enter the next level of arguments, let us
assume 
that detector $B$ makes its measurement first~\cite{first}, in the time
interval 
$[t_B,t_B+\triangle t_B], \triangle t_B>0$ and, subsequently, $A$ proceeds
in 
the time interval $[t_A,t_A+\triangle t_A], \triangle t_A>0$, such that 
$t_A > t_B+\triangle t_B$ holds and the assumption that the measurements
are 
spacelike translates into $c\, (t_A+\triangle t_A-t_B) <
|\vec{x}_A-\vec{x}_B|$.
Assume the measurement result of detector $B$ is $b^k$, $k=\pm 1,...,\pm
n$.
This transforms the state vector $|\Psi_S\rangle$ according to
\begin{equation} \label{transs}
|\Psi_S\rangle \to |\Psi_B^k\rangle
\end{equation}
resulting in new probabilities for $A$, namely
\begin{equation} \label{transp}
p_A^{\pm i} \to p_A^{\pm i,k} = \left| \langle \Psi_A^{\pm i} | \Psi_B^k 
\rangle \right|^2 ~~{\rm for}~~ i=1,..,n;\ k=\pm 1,...,\pm n\, .
\end{equation}
To deliver consistent results, $A$ has to act according to the result of
(\ref{transs}) at (or before) some time $t\le t_A+\triangle t_A$.
Experimentally 
it has been confirmed \cite{Aspect,Tapster} that $A$ does so indeed. To
proceed with an ontological description of QT which attributes reality 
to the state vector (before as well as after measurement), we have to
demand that in coordinate space the wave function collapse
\begin{equation} \label{transw}
\langle \{ x\} |\Psi_S\rangle = \psi_S (\{ x\} ) \to 
\psi_B^k (\{ x\}) = \langle \{ x\} |\Psi_B^k\rangle 
\end{equation}
happens for $x=(ct,\vec{x})$ on a spacelike hypersurface. For simplicity 
we consider in the following only flat planes as hypersurfaces. 
In~(\ref{transw}) the symbolic notation $\{ x\}$ indicates that 
coordinates of several degrees of freedom may be involved. We assume that 
the collapse hypersurface is the same for each degree of freedom.

It is believed that there will be no change in observing the probabilities
(\ref{transp}) when the distance $|\vec{x}_A-\vec{x}_B|$ 
becomes arbitrarily large, e.g. covers astrophysical dimensions. We are 
therefore tempted \cite{tempted} to regard the rest frame of a detector 
as preferred and to conjecture that the transformation  (\ref{transw}) 
happens on the instantaneous hyperplane of this frame, {\it i.e.} is
propagated by $B$ at infinite speed at some sharp time 
$\overline{t}_B\in [t_B,t_B+\triangle t_B]$. The requirement of a 
{\it sharp decision time} $\overline{t}_B$ is necessary to get 
consistent results for the general spacelike situation (\ref{spacelike}),
{\it i.e.} after relaxing our present condition $t_A>t_B+\triangle t_B$. 
Of course, the detector should need a finite response time 
$(\overline{t}_B-t_B)>0$ for getting to its decision and, afterwards,
another $(t_B+\triangle t_B - \overline{t}_B)>0$ for recording the result, 
such that only after time $t_B+\triangle t_B$ it is ready to work 
on its next decision process. In the same way the detector $A$ is supposed
to make its decision at time $\overline{t}_A$. By locating the decisions 
at sharp times and assuming that some continuous stochastic process is 
involved in selecting them, it follows that 
\begin{equation} \label{torder}
 {\rm either}~~ \overline{t}_B < \overline{t}_A ~~{\rm or}~~
                \overline{t}_A < \overline{t}_B\, , 
\end{equation}
whereas the likelihood for $\overline{t}_A=\overline{t}_B$ is zero. In
conclusion, when both detectors are at rest an instantaneous transformation 
(\ref{transw}) does ensure consistent measurements.

For a moving observer the time ordering (\ref{torder}) may become
interchanged. This does not yet imply an inconsistency, as we have 
defined the system of the detector which makes the decision as preferred. 
But the concept of instantaneous state vector reduction becomes
questionable. 
Let us consider detectors $A$ and $B$ 
which move with constant velocity with respect to one another,  
denote the coordinates of the corresponding rest frames by $(x^{\alpha})$ 
and $(x'^{\, \alpha})$, respectively, and locate the detectors at the
origins: $A$ at $\vec{x}_1=0$ and $B$ at $\vec{x}^{\, '}_2=0$. 
Without restricting generality, we can choose 
the orientation of the frames parallel and such that the relative velocity 
is along the $\hat{e}_1$-axis: $\vec{v}=v \hat{e}_1$. With suitable 
definitions of the time and $\hat{e}_1$-axis zero-points, coordinates in
the 
two frames are related by the Lorentz-Poincar\'e transformations:
\begin{eqnarray}
 x'^{\, 0} = +x^0 \cosh (\zeta ) - x^1\, \sinh (\zeta ) , \label{LT1} \\
 x'^{\, 1} = -x^0 \sinh (\zeta ) + x^1\, \cosh (\zeta ) , \label{LT2} \\
 x'^{\, i} = x^i + x^i_{2,0},\ \ i=2,3. \label{LT3}
\end{eqnarray}
Here $\zeta$ is the rapidity variable defined through $\tanh (\zeta )=
\beta=v/c$. The position of $B$ in the frame of $A$ is given by
\begin{equation} \label{xB}
 \vec{x}_B = - \vec{x}_{B0} + \tanh (\zeta)\, x^0\, \hat{e}_1, 
 ~~{\rm where}~~ \vec{x}_{B0} = (0,x^2_{B0},x^3_{B0})\, , 
\end{equation}
in particular $x^1_B = x^0 \tanh (\zeta) $. Assume that $A$ initiates an 
instantaneous transformation (\ref{transw}) at its time $x^0_{A1}$. Then 
$B$ can be aware of the changed probabilities (\ref{transp}) only at or
after its time
\begin{equation} \label{timep}
x'^{\, 0}_{B1} = x^0_{A1}\, \cosh (\zeta) - x^1_B (x^0_{A1})\, \sinh
(\zeta) 
= x^0_{A1}\, { \cosh^2 (\zeta) - \sinh^2 (\zeta) \over
 \cosh (\zeta)} = {x^0_{A1} \over \cosh (\zeta )}\, . 
\end{equation}
The inverse transformations of (\ref{LT1}), (\ref{LT2}) are
$x^0 = x'^{\, 0} \cosh (\zeta) + x'^{\, 1} \sinh (\zeta)$,
$x^1 = x'^{\, 0} \sinh (\zeta) + x'^{\, 1} \cosh (\zeta)$
and the position of $A$ in the frame of $B$ is given by
$\vec{x}^{\, '}_A = \vec{x}_{B0} 
               - \tanh (\zeta)\, x'^{\, 0}\, \hat{e}_1$,
in particular $x'^{\, 1}_A = - x'^{\, 0} \tanh (\zeta) $. 
Therefore, when $B$ initiates at its time $x'^{\, 0}_{B2}$ an instantaneous
collapse, then $A$ can be aware of the result only at or after its time
\begin{equation} \label{time}
 x^0_{A2} = x'^{\, 0}_{B2} \cosh (\zeta) + x'^{\, 1}_A\, \sinh (\zeta) = 
 x'^{\, 0}_{B2}\, {\cosh^2 (\zeta) - \sinh^2 (\zeta) \over \cosh (\zeta)}
 = {x'^{\, 0}_{B2} \over \cosh (\zeta )}\, . 
\end{equation}
The consistency condition to be fulfilled is
\begin{equation} \label{consistency}
{\rm sign}\, (x^0_{A2}-x^0_{A1})\ =\ {\rm sign}\, (x'^{\, 0}_{B2}-x'^{\,
0}_{B1})\, .
\end{equation}
Inconsistent is an arrangement like
\begin{equation} \label{incon}
 x^0_{A2} > x^0_{A1} ~~{\rm and}~~ x'^{\, 0}_{B1} > x'^{\, 0}_{B2}\, , 
\end{equation}
which translates as follows: Using (\ref{time}), (\ref{incon}) 
$$  {x'^{\, 0}_{B2} \over \cosh (\zeta )} = x^0_{A2} > x^0_{A1} $$
and using (\ref{timep}), (\ref{incon})
$$ {x^0_{A1} \over \cosh^2 (\zeta )} = {x'^{\, 0}_{B1} \over \cosh (\zeta
)}
   > {x'^{\, 0}_{B2} \over \cosh (\zeta )}\, .$$
Therefore, a solution of (\ref{incon}) is found for 
\begin{equation} \label{incons}
{x^0_{A1} \over \cosh^2 (\zeta )} > x^0_{A1} 
\Longleftrightarrow x^0_{A1} < 0\, .
\end{equation}
The inconsistency (\ref{incon}) exists when the detectors are approaching
one 
another. Independently of the sign of the rapidity $\zeta$ this happens
for 
negative time, due to our particular choice of coordinates systems. The 
arrangement depicted in figure~1 is for $\zeta >0$ and as seen in the
rest-frame of $A$. A collapse initiated by $A$ at the space-time point 
$A1$ reaches the $B$ world-line at $B1(inst)$ and a collapse initiated by 
$B$ at $B2$ reaches the $A$ world-line at $A2(inst)$.
The contradictory time ordering comes from the fact that the instantaneous 
hyperplane of one inertial frame is not instantaneous in another which
moves with respect to the first. Therefore, we may well admit all 
spacelike planes right away. To avoid the inconsistency, we have to demand 
that the collapse plane of $A$ cuts through the $B$ world-line at a time 
\begin{equation} \label{cons}
 x^0_{B1} < x^0_{B1_{\max}} = x^0_{A1}\, \cosh (\zeta )
\end{equation}
and, symmetrically, that the collapse plane of $B$ cuts through the  
$A$ world-line at
\begin{equation} \label{consp}
 x'^{\, 0}_{A2} < x'^{\, 0}_{A2_{\max}} = x'^{\, 0}_{B2}\, \cosh (\zeta
)\,.
\end{equation}
Here, by definition, $x^0_{B1_{\max}}$ and $x^0_{A2_{\max}}$ are the
largest consistent bounds on the corresponding arrival times for given 
$\zeta$. The $\cosh (\zeta )$ factor follows by exploiting the
symmetry (due to the coordinate definitions) and from the observation
$x'^{\, 0}_{B2}=x'^{\, 0}_{B1_{max}}$, $x^0_{A1}=x^0_{A2_{\max}}$. For
$x^0_{A1} < 0$ we find $x^0_{B1_{\max}} < x^0_{A1}$ and, similarly, 
$x'^{\, 0}_{B2} < 0 \Rightarrow x'^{\, 0}_{A2_{\max}} < x'^{\, 0}_{B2}$,
{\it i.e.} the decision may have to propagate backward in time. This
does not necessarily imply contradictions. Indeed, we had already noticed
that propagation of the collapse process ought to be spacelike and that
comparison of times makes then little sense, because
Lorentz transformations allow to change the time order. On the other
hand, a timelike propagation backward in time would be inconsistent: The 
collapse plane would miss the signal world-lines which connect source
and detectors as indicated in figure~1. Let us determine the collapse 
plane corresponding to $x^0_{B1_{\max}}$ of equation~(\ref{cons}). The 
$x^1$-coordinate of the position of $B$ at time $x^0_{B1_{\max}}$ is 
given by
\begin{equation} \label{pos}
x^1_{B1_{\max}} = \tanh (\zeta)\, x^0_{B1_{\max}} = x^0_{A1}\, \sinh
(\zeta)\, .
\end{equation}
It follows that for finite $\zeta$ the line
$$ (x^0_{A1}, x^1_{A1}=0) \to (x^0_{B1_{\max}}, x^1_{B1_{\max}}) $$
is still spacelike:
$$ (x^0_{A1} - x^0_{B1_{\max}})^2 - (x^1_{A1} - x^1_{B1_{\max}})^2 =
(x^0_{A1})^2\, \{ [1-\cosh (\zeta)]^2 - \sinh^2 (\zeta) \} =
(x^0_{A1})^2\, 2\, [1-\cosh (\zeta)] < 0\, . $$
As detector $A$ does not know about the rapidity of $B$ (or vice versa). 
We have to find a collapse plane which ensures (\ref{cons}) and
(\ref{consp}) for all finite $\zeta$.
From (\ref{cons}) and (\ref{pos}) we notice
$$ \lim_{\zeta\to\infty} {x^0_{B1_{\max}}-x^0_{A1} \over  
x^1_{B1_{\max}} - x^1_{A1} } = \lim_{\zeta\to\infty} { \cosh(\zeta) -1 
\over \sinh (\zeta)} = 1\, . $$
It follows that the {\it only} collapse hypersurface, which may ensure 
consistency for all $\zeta$, is the blc, originating 
at the space-time point $A1$. Similarly, the blc originating at
$B2$ is the only potentially consistent collapse hypersurface when
detector $B$ makes the measurement. In summary, we have shown:
{\it Either the collapse hypersurface is the blc or a 
consistent collapse hypersurface does not exist at all.}
In the following we pursue the blc scenario in further details. In 
particular we address a number of consistency concerns and indicate 
how to overcome them.

In figure~1 we have drawn the light-cones passing through the origin as 
well as the relevant part of the blc originating at $A1$ and $B2$. The blc
of $A1$ hits the world-line of $B$ at $B1(blc)$ and the blc of $B2$ the
world-line of $A$ at $A2(blc)$. Of particular importance are the points
$SA$ 
and $SB$. At $SA$ the $A1$ blc cuts through the $S$--$B2$ line after the 
signal left $S$, but before it reaches $B$ at $B2$. Similarly, at $SB$ 
the $B2$ blc cuts through the $S$--$A1$ line in the
proper way. This implies that $B$ will act correctly at $B2$ when $A$ 
makes the collapse decision at $A1$ and, vice versa, $A$ will act 
correctly at $A1$ when $B$ makes the collapse decision at $B2$. This 
assumes an order of measurements which, as has been understood by now, 
does not necessarily agree with the time ordering of the two 
measurements in the inertial frame of figure~1. We shall come back to 
this point, after discussing other issues.

To be definite, let us assume that $B$ actually makes the collapse decision
at $B2$. The continuous time evolution of the signal propagating towards
$A1$ 
is then interrupted by the discontinuous transformation (\ref{transw})
which 
defines new initial conditions in the neighborhood of the point $SB$
indicated 
in figure~1. Subsequently, causal propagation towards $A1$ continues. A
minimal 
requirement on any ontological formulation of QT appears to demand that
each 
detector will, locally, face only well-defined signals. From the viewpoint
of 
a detector the process is then simple: It acts on whatever signal comes in.

Therefore, when detector $A$ performs now its measurement at $A1$, this
cannot
be allowed to reset at $SA$ the initial conditions of the signal
approaching 
$B$ at $B2$. The explanation at hand is that the collapse initiated by $B$
at 
$B2$ has reduced the quantum correlation to a purely classical one. The
result 
of the transformation (\ref{transw}) has to factorize in the form
\begin{equation} \label{reduce}
|\Psi^k_B\rangle = |\psi_B^k\rangle\, |\psi_A^k\rangle
\end{equation}
such that $\langle x | \psi_A^k\rangle$ propagates along the $SB$--$A_1$
line
and the detector $A$ will only act on $|\psi_A^k\rangle$. It is instructive
to 
write down the thus obtained time development of the
wave function using the coordinates of figure~1 ({\it i.e.} the rest
frame of $A$). We use $x$ to characterize the localization behavior along
$S$--$A1$ and correspondingly $y$ along $S$--$B2$. For $x^0=y^0=ct$ we
find:
\begin{equation} \label{wfct}
\langle y, x | \Psi \rangle = \cases{
\langle y, x | \Psi_S \rangle ~~{\rm for}~~ t< t(SB),\cr
\langle y | \psi_B\rangle\, \langle x | \psi_A^k\rangle  
   ~~{\rm for}~~ t(SB) < t < t(B2),\cr
\langle y | \psi_B^k\rangle\, \langle x | \psi_A^k\rangle  
   ~~{\rm for}~~ t(B2) < t < t(A1),\cr
\langle y | \psi_B^k\rangle\, \langle x | \psi_A^l\rangle  
   ~~{\rm for}~~ t(A1) < t.\cr}
\end{equation}
Discontinuous jumps occur at $t(SB)$, $t(B2)$ and $t(A1)$. In between we
have continuous, causal time evolution. For the time range 
$t(SB)<t<t(B2)$ we use
the symbol $|\psi_B\rangle$ instead of $|\psi^k_B\rangle$, because the
$y$-coordinate has not yet passed the collapse blc. The ansatz
$|\psi_B\rangle \sim\, _A\langle \psi_A^k | \Psi_S\rangle$ gives consistent 
results and the factor is fixed by requesting proper normalization. A
singlet state which decays at $S$ into two distinguishable spin 1/2 
particles may serve as an example. Then~\cite{Bo93}
\begin{equation} \label{singlet}
 |\Psi_S\rangle = 2^{-{1\over 2}}\, \{ |-\rangle_B\, |+\rangle_A
                - |+\rangle_B\, |-\rangle_A \} 
\end{equation}
where the subscripts $A$, $B$ denote the distinguishable particles as well
as 
the branches on which they move towards the corresponding detectors. The
wave 
functions of equation (\ref{wfct}) are now
$$ \langle y, x | \Psi_S \rangle = 2^{-{1\over 2}}\, \{ 
           \langle y | -\rangle_B\ \langle x | +\rangle_A 
         - \langle y | +\rangle_B\ \langle x | -\rangle_A  \} $$
$$ \langle y | \psi_B\rangle\, \langle x | \psi_A^k\rangle  =
2^{-{1\over 2}}\, \{ \langle y |-\rangle_B - \langle y |+\rangle_B
                  \}\, \langle x | a^{-k} \rangle_A $$
$$\langle y|\psi_B^k\rangle\, \langle x|\psi_A^k\rangle = 
  \langle y | b^k \rangle_B\ \langle x | a^{-k} \rangle_A $$
$$\langle y|\psi_B^k\rangle\, \langle x|\psi_A^l\rangle = 
  \langle y | b^k \rangle_B\ \langle x | a^l \rangle_A $$
where $b^k$, $a^{-k}$ and $a^l$ indicate measurement results as defined 
by equation~(\ref{alternatives}). Let us stress that our reduction 
scenario does not rely on any particular choice of coordinates, because
the blc is a geometric object which has the desireable property to stay 
invariant under Lorentz transformations. The point $SB$ is as physical 
as the points $B2$ and $A1$ are. The choice of coordinates is secondary. 
Just one more example, in the rest frame of $B$ (indicated by primed 
coordinates and, as before, $x'^{\, 0}=y'^{\, 0}=ct'$) we get, 
for instance,
$$ \langle y', x' | \Psi \rangle = \langle y' | \psi_B\rangle\, 
\langle x' | \psi_A^l\rangle ~~{\rm for}~~ t'(A1)<t'<t'(B2)\, .$$
This may be compared with the previous result for $t(B2)<t<t(A1)$,
just re-iterating the fact that an absolute time does not exist. 
(The values corresponding to figure~1 are $c\,t_{A1}=-1$,
$c\,t_{B2}=-1.052$
versus $c\,t'_{A1}=-1.128$, $c\,t'_{B2}=-0.933$.)

The case where the signals propagate at the speed of light towards the 
detectors may be perceived as another difficulty. It is undesirable 
that the collapse surface cuts through the source. Fortunately, the 
problem does not really exist, due to the finite detector response 
times.  For lightlike signals, and emphasizing now the response time 
of detector $B$, this is illustrated for the source $S2$ in figure~1.

Let us return to the ordering problem for our measurements. QT does not 
specify such an order and the observable correlations do not depend on it,
as spacelike operators ${\cal A}$ and ${\cal B}$ commute. In our context
a specific order is needed to arrive at a complete space-time picture
for the wave function. A surprisingly simple and explicitly Lorentz 
invariant solution exists. One may define the order by comparing proper
times (of the detectors) $\tau_A$ and $\tau_B$, which the signals need to 
propagate from the source (creating the quantum correlation) to the 
detectors, $\tau_A$ from $S$ to $A1$ and $\tau_B$ from $S$ to $B2$. 
Corresponding to figure~1: $c\,\tau_A=0.8$ and $c\,\tau_B=0.664$. 
Randomness should be involved when those times overlap within the
uncertainties set by the detector response times $\triangle\tau _A$ and
$\triangle\tau_B$. By reasons already discussed, the reduction time itself,
more precisely the corresponding blc, should be sharp.
  
Whatever the rule for the order of detector decisions is, it will not 
imply deviations from {\it standard} QT. However, an 
ontological formulation of QT invites to transgress beyond standard 
QT. Naturally, it allows to address problems, eventually with observable 
consequences, which hardly can be properly addressed otherwise. Namely, 
the dynamical process of QT has since long \cite{vonNeumann} been 
perceived as continuous, causal time evolution, interrupted by
discontinuous
jumps, called measurements. Measurements are achieved by applying detectors
to wave functions. The question whether continuous, causal time evolution
of an enlarged wave function, now including the detectors and eventually
the environment beyond, will yield consistent results has been 
an issue of controversial debate \cite{Wheeler,Stamatescu}. Our ontological
QT formulation adds to this. Within its framework the answer becomes 
explicit: Continuous time evolution and jumps are distinct. Including the
detectors in the wave function will, up to minor notational
re-arrangements,
not change the discontinuous reduction (\ref{transs}),(\ref{reduce}) and
there is no way that inclusion of detector $B$ in the wave function can 
lead to a causal explanation of the spontaneous change at $SB$. Actually,
for
consistency the detectors should have been included, as the states 
$|\psi_A\rangle$ and $|\psi_B\rangle$ 
will obviously be mixed-up with the corresponding measurement
devices. Whether our explicit rules reflect reality or not may remain
undecidable. However, as long as they cannot be excluded, either by 
reasons of inconsistency or by experimental facts, the often proclaimed
consistency of QT measurements with the time evolution of an enlarged
system remains hypothetical and, actually, unlikely.

The theory of reduction (jumps) exist only to the extent that QT makes
statistical predictions for the case when a state vector reduction actually
happens. In addition, there are good reasons to believe that the
stochastic
character of these predictions is of fundamental nature. What is missing 
in QT, and becomes only in the ontological formulation an obvious 
incompleteness, is an understanding of the microscopic properties of matter 
which are responsibly for the ability of detectors~\cite{detector} to make 
collapse decisions. We only know that reduction 
happens before macroscopic superpositions of detectors emerge. But it is 
certainly not satisfactory to attribute the process to emergent, ad-hoc 
properties of
macroscopic matter. Instead, a search for hereto overlooked new,
fundamental
properties of microscopic matter seems to be legitimate. The framework 
presented in this paper allows rather naturally to make 
heuristic-phenomenological assumptions and work along 
this line is presented in ref.~\cite{Berg}.
 
In conclusion, the quantum state vector may provide a glimpse of reality. 
A reality which is clearly very different from the traditional, realistic
vision of nature~\cite{EPR}. Essentially, it was the achievement of 
Bell~\cite{Bell} that the latter is, on experimental grounds, now 
conclusively ruled out. It seems to be time to move forward, to
develop a new view of reality on the basis of QT.
The discussion of this paper has been limited to the example depicted 
in figure~1. Hence, the derived constraints appear to be necessary
ingredients of a broader, ontological understanding of QT. Whether 
a general formulation can be worked out consistently remains to be seen,
although the author is optimistic. After all, assuming the existence
of reality in form of the state vector provides strong guidance.
\medskip

\noindent
{\bf Acknowledgement:} I would like to thank Wolfgang Beirl 
for his interest and many useful discussions.


\begin{thebibliography}{12}
\bibitem{first} At this point time ordering is defined with respect 
                to the common rest frame of both detectors, such 
                that $t_B=t_{B2}<t_{A1}=t_A$ in figure~1.
\bibitem{tempted} Naively, ``instantaneous'' appears to be a natural
                  choice. It turns out to be wrong, but is instructive 
                  anyway.
\bibitem{detector} Here the word detector refers to any conglomerate
of matter which has the ability to make collapse decisions. A perception
that this ability is limited to man-made detectors or even human 
observations appears unphysical. The physical assumption is that time
evolution of nature (wave function of the universe) proceeds according
to two laws: causal time evolution and reduction. It is not well
understood when the latter happens. This makes astrophysical approaches,
essentially invoking only the first law, quite questionable.
\end{thebibliography}
\end{document}